\documentclass{basinew}
\usepackage{txfonts}
\usepackage{graphics}
\usepackage{chicago}
\usepackage{psfig}
\usepackage{subfigure}

\begin{document}

\title{ A software baseband receiver for pulsar astronomy at GMRT}
\author[B. C. Joshi and Sunil Ramakrishna]{B. C. Joshi$^{1}$\thanks{e-mail:bcj@ncra.tifr.res.in} and ~~Sunil Ramakrishna$^{2,3}$ \\
$^{1}$National Centre for Radio Astrophysics (TIFR), Pune 411 007, India \\
$^{2}$International Institute of Information Technology (I2IT), Pune - 411 057, India \\
$^{3}$Intoto Software(I) Pvt Ltd., Hyderabad 500 082, India  \\}
\date{Received 19 05 2006; accepted 07 09 2006}
\maketitle
\label{firstpage}

\begin{abstract}
A variety of pulsar studies, ranging from high precision astrometry 
to tests for theories of gravity, require high time resolution data. 
Few such observations at more than two frequencies below 1 GHz are 
available. Giant Meterwave Radio Telescope (GMRT) 
has the unique capability to provide such multi-frequency pulsar 
data at low observation frequencies, but the quality and 
time resolution of pulsar radio signals is degraded due to 
dispersion in the inter-stellar medium at these frequencies. 
Such degradation is usually taken care of by employing 
specialized digital hardware, which implement coherent 
dedispersion algorithm. 
In recent years, 
a new alternative is provided by the availability of 
cheap computer hardware. In this approach, the required 
signal processing is implemented in 
software using commercially off-the-shelf available computing 
hardware. This makes such a receiver flexible and upgradeable 
unlike a hardware implementation. The 
salient features and the modes of operation of a high time resolution pulsar 
instrument for GMRT 
based on this approach is described in this paper. The capability 
of the instrument is demonstrated by illustrations of test 
observations. We have obtained the average profile of PSR B1937+21 at 
235 MHz for the first time and this profile indicates a scattering 
timescale of about 300 $\mu$s. 
Lastly, the possible future extensions of this concept are discussed. 
\end{abstract}


\section{\bf Introduction}

\label{sec1}

High time resolution data are required in a variety 
of pulsar studies. Useful constraints on the location 
and size of emission region as well as the emission 
mechanism can be obtained by studies of narrow intense 
highly polarized Giant Pulses (GPs), such as those seen 
in PSRs B0531+21, B1937+21 and a few other pulsars 
\shortcite{sb95,kt00,hkwe03}. Further constraints on emission 
mechanism for pulsars are placed by high time resolution 
observations of micro-pulses, observed in pulsars such 
as PSRs B1133+16, B0950+08 and J0437$-$4715 
\shortcite{jak+98,pbc+02,kjv02}. Polarization 
observations for millisecond pulsars (MSPs) are being 
carried out increasingly with high time 
resolution instruments \shortcite{jak+98,stc99,ovhb04}, 
primarily due to the short periods of these pulsars. 
These observations, which  appear to show a flat polarization 
angle swing for MSPs as opposed to an S-Shaped curve 
for most normal pulsars, are needed to constrain a 
probably more complex magnetospheric physics of these 
stars.  Such observations are also required to discriminate 
between the models explaining orthogonal polarization 
mode changes \shortcite{ms98}. High time resolution 
observations also provide high precision astrometric 
measurements as illustrated by the distance measurement 
of PSR J0437$-$4715 system using annual-orbital parallax 
\shortcite{vbb+01}. In addition, such data are 
useful in experiments involving tests for general theory of 
relativity \shortcite{tw89,kra98,vbb+01,wt02,lbk+04} and 
the detection of primordial gravitational wave background 
\shortcite{fb90}. Few such observations at more than two 
frequencies, simultaneous or otherwise,  are available, 
particularly for observation 
frequencies below 1 GHz.

Giant Meterwave Radio Telescope (GMRT) 
has the unique capability to provide 
such multi-frequency high quality pulsar data at observation frequencies 
ranging from 150 MHz to 1420 MHz \shortcite{sak+91}. 
As pulsars exhibit a steep 
spectra with typical spectral index of -1.8 
\shortcite{mkk+00}, the  
signal to noise ratio (SNR) of pulsar data increases 
with decreasing observation 
frequencies upto 235 MHz at GMRT. 
Large observation bandwidths (typically 300 MHz to 1 GHz) 
are employed at high frequencies 
as the data quality 
improves with the square root of observing bandwidth. 
Although 
the observation bandwidths are limited at 
GMRT due to smaller front-end design bandwidths as 
well as a much larger Radio Frequency Interference (RFI), 
GMRT is a multi-element telescope consisting of thirty 45-m diameter 
dishes, which can be phased to provide a single 
dish with effective collecting area ranging from 12000 
m$^2$ $-$ 30000 m$^2$. Thus, GMRT has the 
unique capability to trade-off number of antennas with 
observation bandwidth. 
Consequently, the data quality is comparable or better than most 
single dish telescopes operating at high frequencies 
despite the limited bandwidth.

The time resolution and quality of pulsar data is 
however limited due to dispersion in the interstellar 
medium (ISM), particularly at low frequencies of operation at GMRT. 
The effect of dispersion in ISM can be mitigated using 
a large filterbank. For example, the standard pulsar 
hardware at GMRT 
employs a 256 channel digital filterbank. Although 
this reduces the dispersion smear to about 114 $\mu$s 
per frequency channel across a 16 MHz bandpass 
for a pulsar with a Dispersion Measure ( DM - the integrated 
column electron density along the line of sight to the pulsar) 
of 50 pc$\,$cm$^{-3}$ at a frequency of 610 MHz, the corresponding 
smear are very large at lower frequencies used at GMRT (750 $\mu$s, 2 ms 
and 8 ms for 325, 235 and 150 MHz respectively). 
This dispersion of pulsar signal due to ISM can be 
eliminated using a receiver which implements 
coherent dedispersion algorithms \shortcite{hr75}. 
As the pulsar hardware currently in use at GMRT does not provide 
data with sufficient time resolution, particularly for 
high DM short period pulsars at the lower operating 
frequencies, a new pulsar instrument with coherent dedispersion 
capability was required at GMRT to 
provide such data and to fill this 
gap in GMRT's capabilities. 

Traditionally, the coherent dedispersion algorithms 
were employed in specialized hardware using Digital 
Signal Processing (DSP) or Field Programmable Gate 
Array (FPGA) chips. Although these designs are low cost, 
the functionality of such hardware 
is frozen at the time of design, making them inflexible to 
changing demands of pulsar astronomy. A new alternative 
is provided by the availability of cheap computer 
hardware in recent years. In this 
approach, the required signal processing is implemented 
in modular and portable software using commercially 
off-the-shelf available computing 
hardware. Since the functionality of the receiver is 
defined in software, such a receiver is more flexible. 
One of the first such pulsar backends was Coherent Baseband 
Receiver for Astronomy (COBRA) and similar backends have been made 
operational at Westerbok, Arecibo, Parkes and Green Bank 
Observatories recently \shortcite{jlk+03,jlk03}. A new  
pulsar instrument for GMRT, based on this approach, 
is  described in this paper. The salient features of  
this instrument are described in Section \ref{sec2}. 
The results of test observations, demonstrating the 
high time resolution capability of the instrument, are 
discussed in Section \ref{sec3}. Finally, the future 
development of this approach is outlined in Section 
\ref{sec4}.

\section{\bf Pulsar Software Baseband Receiver} 

\label{sec2}

\begin{center}
\begin{figure}
\psfig{file=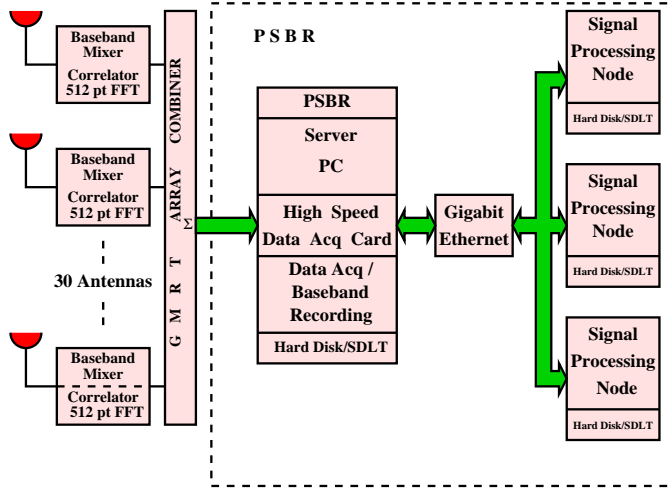,width=3.5in}
 \caption{The schematic diagram of PSBR}
\label{fig1}
\end{figure}
\end{center}

Pulsar Software Baseband Receiver (PSBR) implements the signal processing, 
required for obtaining high time resolution high SNR  
pulsar data, using portable, open-system, flexible and upgradeable software. 
PSBR consists of a commodity high speed data acquisition card and a Beowulf cluster 
of 4 off-the-shelf commodity personal computers (PC) connected 
by a Gigabit Ethernet switch as shown in Figure \ref{fig1}. The current 
configuration of PSBR uses PCs with single 2.3 GHz processor 
running open-source Linux, 
1 GBytes random access memory (RAM), 200 GBytes disks and a 
motherboard with Peripheral Connect Interface (PCI) bus. 
A high speed acquisition card that supports a 
32K words First In First Out (FIFO) buffer and data transfer 
to user RAM using scatter-gather Direct Memory Access (DMA) 
at 80 MBytes/s sustained data transfer rates 
on a typical PCI bus, is used to acquire digitized raw data.

\begin{center}
\begin{figure}
\psfig{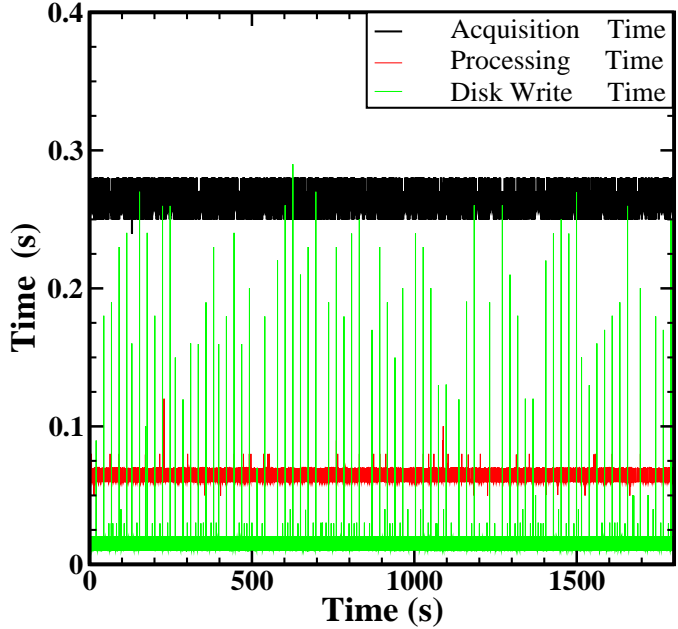}
 \caption{The figure shows the execution times for acquisition, 
          processing and disk write modules for 7200 acquisition 
          buffers of 16 MBytes each as a function of 
          observation time for a 30 minute observation in 
          baseband recording mode of PSBR. A full bandwidth 
          acquisition of 16 MBytes buffer typically takes 0.25 
          s as shown by dark points. The processing and disk 
          write take less than 0.09 s on average, thus allowing 
          sustained data capture. Note that disk write execution 
          times show 
          spikes, necessitating provision 
          of adequate cache buffers to avoid data loss}
\label{fig2}
\end{figure}
\end{center}

The existing analog 
and digital data pipeline of GMRT is used to obtain baseband 
data for the two polarizations of the received radio frequency signal. 
A 512 point  Fast Fourier Transform 
(FFT), followed by phase compensation for each antenna \shortcite{sir00}, 
is carried out by the 
GMRT correlator \shortcite{sdt+95} 
in this pipeline and these phased voltages from  
different antennas are added using 
GMRT Array combiner \shortcite{d95}. 
These data are acquired in PSBR at a rate of 64 MBytes/s (corresponding to 
a bandwidth of 16 MHz) using the high speed acquisition board 
mounted in the server PC, which farms 
out data by either a Transmission Control Protocol (TCP) 
or Message passing interface (MPI) over the Gigabit switch 
to signal processing PCs. 

\begin{center}
\begin{figure}
\psfig{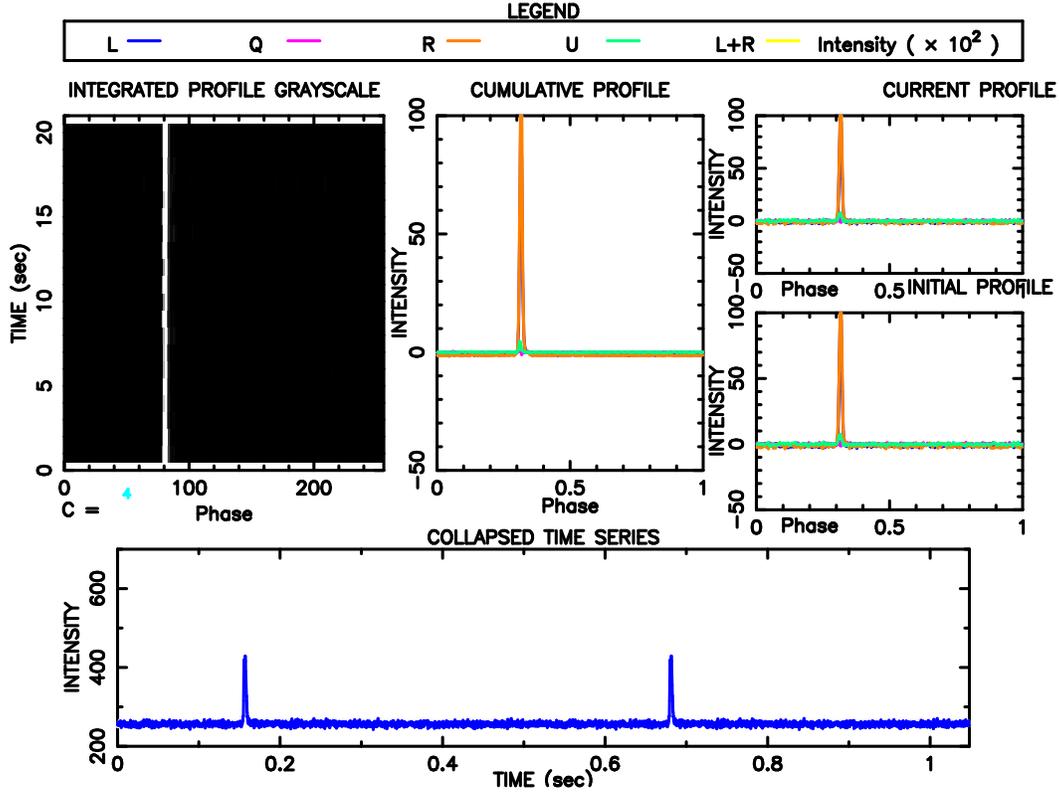}
 \caption{The on-line integrated profiles display in the polarimeter 
          mode of PSBR}
\label{fig3}
\end{figure}
\end{center}

The PSBR has three modes of operations : (1) high time resolution 
baseband recording mode, (2) Polarimeter and Incoherent 
array mode and (3) on-line coherent dedispersion mode. 
The acquired data are recorded to a local hard-disk or
SDLT tape with a sampling time of 250 ns and 8 bit 
precision in the baseband 
recording mode for off-line coherent dedispersion. 
As the data 
acquisition rate is close to the maximum possible sustained 
bandwidth supported by the motherboard used, extensive 
caching is used in the software. The baseband recorder 
program is a multi-threaded program, which carries out 
acquisition, data sequence checks and disk write functions 
concurrently. Each concurrent stage has three to eight 
16 MBytes cache buffers to maintain the continuity of the data. 
Figure \ref{fig2} illustrates the typical time taken by the 
concurrent sections in a 30 minute 4 MHz observation. 
The available observation bandwidth 
is limited by the available motherboard bandwidth 
and disk write rate, which is probably the cause of the 
spiky behaviour in disk write times visible in Figure \ref{fig2}. 
Tests have indicated a stable operation 
for a maximum observation bandwidth of 4 MHz, although 
data for a bandwidth of 8 MHz can also be acquired under ideal 
conditions. This translates to a typical data volume of 
56.25 GBytes per hour for a bandwidth of 4 MHz. The data is 
backed up on an available SDLT tape drive after each observation.

The first mode was designed to make use of the GMRT Array Combiner 
to carry out simultaneous multi-frequency observations 
of pulsars and for observations of pulsars with uncertainties in known DM. 
GMRT Array Combiner provides a facility to selectively mask any of 
256 frequency 
channels provided by the digital filterbank implemented in 
the GMRT correlator. This feature is used to group 
available antennas in two phased array groups operating 
at different observational radio frequency simultaneously. 
Such a configuration can be used with baseband recording mode 
of PSBR to carry out simultaneous multi-frequency observations 
of pulsars. This mode, exploiting the unique multi-element multi-frequency 
nature of GMRT, is very useful for spot DM determinations. 
An example of such observations is illustrated later. The 
estimated value of DM can then be used to get high quality 
folded profiles using the on-line coherent dedispersion mode later.

In the polarimeter mode, the acquisition software carries
out the relevant complex multiplications 
of the two polarized voltages on-line to generate full stokes
parameter time series, which can then be 
integrated upto a desired sampling time (in excess of 
128 $\mu$s). This reduced data is then farmed over the 
Gigabit Ethernet to signal processing PCs using TCP/IP protocols, 
where different software modules carry out on-line
incoherent dedispersion / folding and simultaneous display 
in one of three different formats 
- bandshapes, collapsed integrated profiles and full 256 channel
integrated profiles.
The integrated profiles display, 
shown as an illustration in Figure \ref{fig3}, shows the integrated profile 
for the initial and current sub-integrations, a grey scale plot 
of the dedispersed integrated  pulsar data as a function of pulse phase 
and sub-integration number and a part of the dedispersed time series. 
Similar plots emphasizing the received intensity as a function of 
channel number are shown in the bandshapes and 256 channel
integrated profiles displays. These different on-line displays 
provide an on-line feel of the different aspects of 
data to the observing astronomer and are useful in monitoring 
the quality of observations. The data for full bandwidth 
available from a GMRT sideband are written to a user specified 
disk file or an SDLT tape in the signal processing PC with 
16 bits precision. 
The disk/tape write rate and the data volume depends on the 
integration time selected for this mode and is  
15.26 MBytes per second and 53.6 GBytes per hour respectively 
for the fastest mode which has been tested for stable operation 
(sampling time $-$ 128 $\mu$s ;  observation bandwidth $-$ 16 MHz). 
This mode was designed with a view to provide a backup backend 
in case of a hardware failure of the old GMRT pulsar backends 
and can support all current pulsar observations with GMRT. 

In the on-line coherent dedispersion mode, the baseband data 
for 2 MHz bandwidth 
is dynamically farmed to one of the three signal processing nodes 
using MPI protocols for coherent dedispersion and for 
subsequent display and recording of reduced average 
pulse profiles. The MPI programs treat the entire 
Beowulf cluster as a single 4 processor system and implement 
data acquisition, inter-processor communications and 
signal processing in logically separated, concurrent and 
independent modules. The communication module keeps track 
of the availability of signal processing nodes and the processed data 
and accordingly allocates newly acquired buffers to idle nodes. 
It also collects processed data and integrates these into 
sub-integrations of user specified duration, which are written 
to disk as well as displayed on-line. The signal processing 
module carries out coherent dedispersion by convolving the 
raw data for both the polarizations with a filter, 
which implements a transfer function inverse to that of ISM  
\shortcite{hr75}. This is followed by a phase coherent 
folding of the pulsar time-series. The software is so 
organized that this module can be replaced by another 
module implementing alternative signal processing such 
as a software filterbank making the functionality of the 
receiver flexible. The typical data volume for this mode 
depends on the user specified sub-integration time and the 
desired number of bins across the integrated profile. It is 
typically less than a few MBytes per hour. This mode is 
intended for high precision timing of the pulsars with 
well determined DM on a routine basis.

The data acquisition is
synchronized with the minute pulse from Global Positioning System 
(GPS) for generating time-stamps for all three modes of PSBR. 
The modes can be configured from user specified 
parameter files 
and can be controlled from Telescope control using 
simple commands. 

The signal processing for each of this functionality 
is implemented in software in a modular form so as to allow 
modification / upgradation by replacement with new functional modules. 
The definition of functionality of the receiver in software 
makes it flexible as opposed to earlier hardware implementations. 
This is best illustrated by the three different modes of PSBR, 
which use the same general purpose hardware. 
Thus, the design of the receiver both in its hardware and software 
components resembles the concept of a Software Defined Radio (SDR), 
which is fast becoming popular in industry for similar reasons. 

\begin{figure}
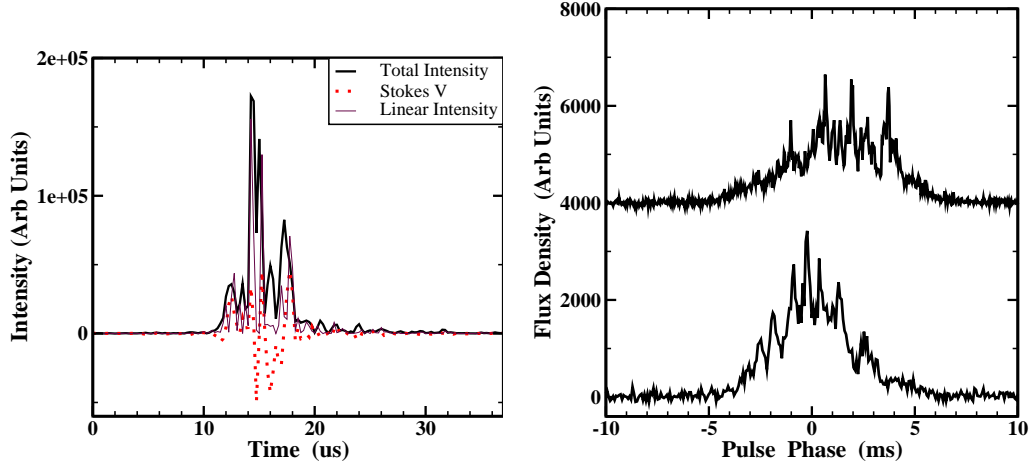

\centering
\subfigure{\mbox{\psfig{file=fig4a.eps,width=2.6in}}}\quad
\subfigure{\mbox{\psfig{file=fig4b.eps,width=2.6in}}}
 \caption{(a) The left plot shows uncalibrated stokes profiles of a 
          GP in PSR B0531+21. 
              The solid line shows the total 
              intensity, the dotted line circular polarization and the grey 
              line total linear polarization
          (b) The right plot shows the coherently dedispersed time 
           series exhibiting micro-pulses for PSR B1133+16 at 610 MHz 
           obtained with PSBR}
\label{fig4}
\end{figure}

\section{\bf Results} 

\label{sec3}

The baseband recorder and the polarimeter mode of PSBR have 
been under test since October 2004 and have given a stable 
performance. The on-line coherent dedispersion mode is being 
tested currently and efforts are on to enhance bandwidth in 
this mode. The main objective of this receiver was to obtain high 
time resolution pulsar data for studies of the kind indicated 
in Section \ref{sec1}. Hence, test observations of 
suitable pulsars were carried out and the results, 
highlighting the capabilities 
of the instrument, are discussed in this section.

A handful of pulsars show narrow micro-second scale intense single pulses, 
called Giant Pulses (GPs),  with 
intensities exceeding the average intensity by a factor of 
100. Coherently dedispersed observations of such pulses are 
required to resolve the structure in these pulses. Such observations provide 
a good test for the high time resolution capabilities of 
PSBR. Figure 4a shows the profile of a 
GP of PSR B0531+21 in data obtained 
at 610 MHz with 250 ns time resolution. The GP profile 
clearly shows the nano-pulses reported earlier \shortcite{hkwe03} 
and is consistent with their observations. Unique high quality data 
for single pulses from this 33 ms pulsar can be obtained using PSBR with GMRT as 
the phased array of GMRT is capable of resolving the  
strong radio emission from the surrounding nebula, thus 
allowing a probe into the weaker radio emission of the pulsar. 
This figure also 
shows uncalibrated linear and circular polarization profile 
suggesting that GPs are highly polarized.

\begin{figure}
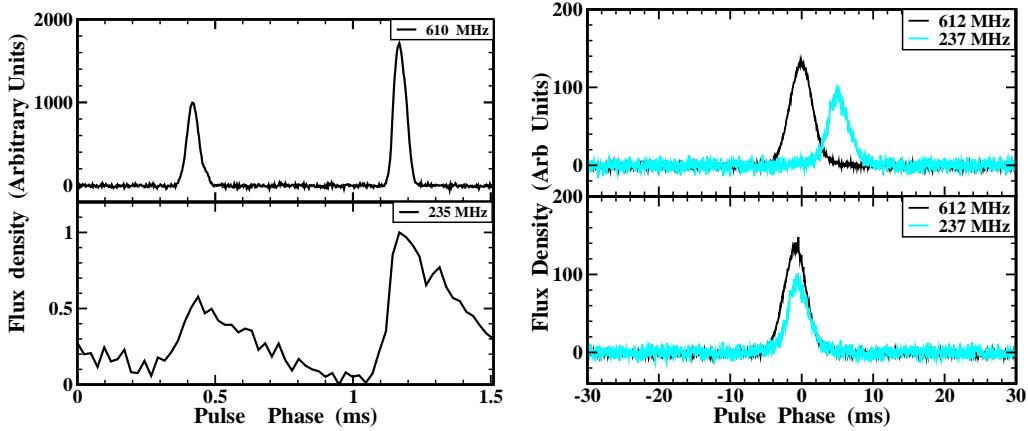

\centering
\subfigure{\mbox{\psfig{file=fig5a.eps,width=2.6in}}}\quad
\subfigure{\mbox{\psfig{file=fig5b.eps,width=2.6in}}}
 \caption{(a) The left plot shows the coherently dedispersed average 
           profiles of MSP PSR B1937+21 at 610 MHz (upper panel) and 
           at 235 MHz (lower panel) obtained with PSBR. 
          (b) The right plot shows the pulse averaged over 700 periods of PSR 
              B1642$-$03 obtained at 613 MHz and 237 MHz simultaneously, 
              where the bins corresponding to $\pm$ 0.1 phase around 
              the pulse are plotted. The upper panel shows the averaged 
              pulse at the two frequencies with an assumed DM of 
              35.665 pc\,cm$^{-3}$, while the lower panel shows that 
              with the estimated DM (35.740 pc\,cm$^{-3}$)}
\label{fig5}
\end{figure}

Many pulsars show structure at micro-second scale in their single pulses, 
which is called micro-structure. Figure 4b shows high resolution 
observations of two single pulses of PSR B1133+16 in data 
obtained at 610 MHz with 500 ns time resolution. The pulses 
clearly show significant spikes, probably with a characteristic periodicity. 
These micro-pulses are consistent with previously reported observations  
\shortcite{pbc+02}. Such studies, particularly as a function of 
frequency of observation, will be useful in constraining 
the pulse emission mechanism. 

PSR B1937+21, 
a millisecond pulsar with the second shortest known 
period (1.5 ms) and a high 
Dispersion Measure (DM $\sim$ 71 pc\,cm$^{-3}$), is another pulsar best 
suited to characterize the high time resolution capability 
of PSBR. The current GMRT backends give only 12 bins across its 
average profile at 610 MHz, which has a dispersion smear of about 2 bins. 
The dispersion smear is larger than the period, when this pulsar 
is observed with the current GMRT backends at 235 MHz. 
The upper plot of Figure 5a shows the coherently 
dedispersed average profile of this pulsar obtained at 610 
MHz with a sampling time of 250 ns 
and folded to 512 bins with an effective bin-size of 3 $\mu$s. 
The corresponding profile at 235 MHz smoothed to 64 bins is shown in the lower 
panel of Figure 5a. 
Whereas the 610 MHz profile is consistent with similar observations at this 
frequency, the 235 MHz profile has been obtained for the first 
time to the best of our knowledge. The pulsar is not 
detected in observations with old backends, which 
provide incoherently dedispersed data. 
Such profiles will be useful to study the 
evolution of average profile components with observing frequency 
for a much larger sample of MSPs available 
now as compared to those reported earlier \shortcite{kll+99}. 
Moreover, PSBR gives the full Stokes profile and this will be useful 
in polarization studies \shortcite{stc99}, particularly as a function 
of observation frequency.

\begin{figure}
\centering
\psfig{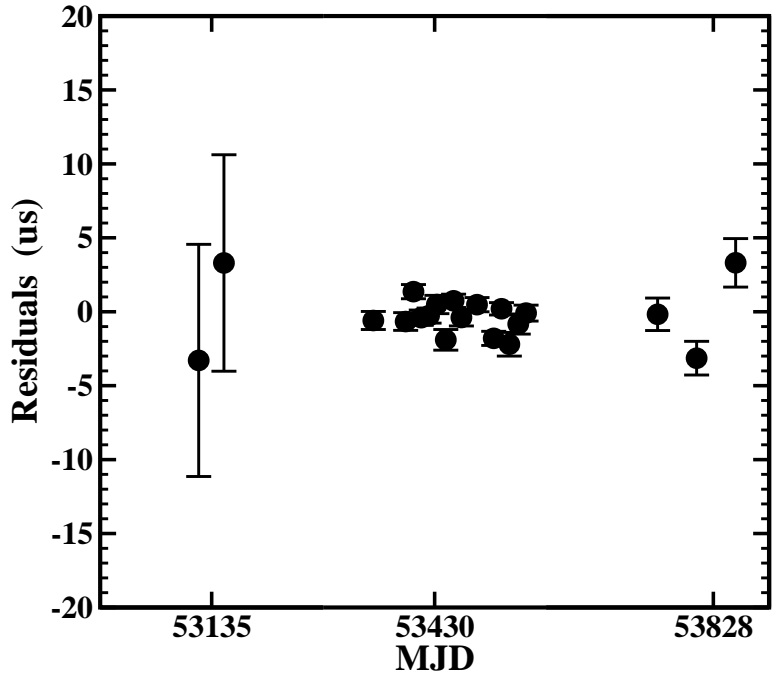}
 \caption{The plot shows TOA residuals for PSR B1937+21 in observations 
          carried out at three epochs, separated by about a year, after 
          fitting a suitable pulsar model independently for each epoch. 
          The MJD for these epochs is indicated along the X-axis. 
          For clarity, the scale along 
          X-axis is expanded around each epoch and is not uniform.
          The measurements over about three hours at each epoch are shown. 
          The data near MJD 53135 represent the  
          measurements carried out in May 2004 with the low 
          resolution backend. The rest of data  
          represent recent measurements using PSBR. 
          The typical errors for the latter are 1 $\mu$s} 
\label{fig6}
\end{figure}

The lower frequency profile in Figure 5a 
shows broadening of the pulse due to scattering in 
ISM. Assuming an exponential scattering 
tail, these two frequency measurements yield a scattering 
time scale of $\sim$ 300 $\mu$s. This estimate, together 
with previous measurements \shortcite{kt00}, indicates 
that scattering towards this pulsar scales as $\nu^{-4}$, 
consistent with turbulent scattering\footnote{Scattering frequency 
index for Kolmogorov density spectrum is $-$4.4. See \shortciteN{r96}}. 
Thus, such multi-frequency profiles will be 
useful probes for studies of scattering in ISM. Although the scatter 
broadening seen for this MSP does substantially reduce the 
time resolution for MSPs at frequencies below 300 MHz, it 
should be noted that the scatter broadening is a function 
of Galactic longitude as well as Galactic latitude. 
Consequently, it is possible to get reasonable time 
resolution for many high latitude MSPs at low radio frequencies, 
particularly for the MSPs in the anti-center direction.

Simultaneous high time resolution multi-frequency observations 
are useful to obtain accurate estimates of DMs of pulsars. 
Figure 5b shows the averaged pulses  
for PSR B1642$-$03 obtained using GMRT Array Combiner and PSBR in two frequency 
simultaneous baseband recording mode (explained in Section 
\ref{sec2}) with GMRT antennas operating at  613 and 237 MHz. The smoothed 
time resolution of the two profiles is 50 $\mu$s and the pulsar was 
observed for 300 s. The difference in arrival 
time of the pulse in the upper panel of Figure 5b 
is due to the error in the assumed 
nominal DM for this pulsar and can be used to estimate 
the required correction to the nominal DM. The lower panel 
of Figure 5b shows the averaged pulse at the two 
frequencies after the 
estimated correction was applied. It is possible to obtain 
high SNR profiles with a time resolution of 10 $\mu$s 
in half an hour to two hour observations 
using GMRT phased array for most pulsars. With this 
time resolution, it is possible to measure DM offsets 
of the order of 0.0001 pc\,cm$^{-3}$, provided the pulsar 
exhibits an average profile with a sharp pulsed feature. Such 
observations  over multiple epochs can also be used to 
study the fluctuations in electron column densities towards 
the pulsar.

As PSBR is likely to play an important role in high precision pulsar 
timing, test observations of PSR B1937+21 to characterize its 
timing accuracy were also carried out. Time of 
Arrival (TOA) data, acquired during these tests, were analyzed 
together with older low time resolution data on this pulsar 
using the pulsar timing package TEMPO to estimate 
the improvement in timing errors for MSPs. Figure \ref{fig6} shows TOA 
residuals obtained with PSBR for this pulsar after a suitable 
pulsar model fit  in comparison with low time resolution data. 
As is evident from the figure, residuals of the order of 1 $\mu$s are 
routinely possible, even at 610 MHz, for short observations. 

\section{\bf Planned Improvements} 

\label{sec4}

As PSBR is implemented in software, a similar receiver 
for the second sideband of GMRT can be commissioned very 
rapidly and such a receiver is being planned. As the two 
sidebands are independent pipelines, this will enhance the 
capability for simultaneous two frequency observations. In 
addition, a new interface card, which will allow the data 
to be acquired from either the incoherent array output or 
the phased array output of GAC under full software control, 
is currently being commissioned. The above development 
will provide a complete backup of the existing pulsar 
backends and further underscores the flexibility of the 
software based instruments.

As has been known since the time COBRA was designed, the bottleneck in 
extending the bandwidth of such a  system without 
multiple copies is the rate ($\sim$ 32 MBytes/s) at which data 
can be transfered to PC RAM through PCI  
bus  \shortcite{jlk+03}. Recently, PC manufacturers adopted a new 
standard for this bus,  known as PCI Express (PCIe), which promises to push 
this bandwidth  to beyond 16 GBytes/s. 
This bus has been adopted enthusiastically by the industry 
and an extension of PSBR using PCIe based data acquisition 
cards is planned.

Lastly, it is desirable to develop a wide-band ($\sim$ 400 MHz) single 
antenna clone of PSBR with new high speed digitizers and 
a Beowulf cluster of PCs, reusing much of the software developed for 
PSBR. Single antenna operation eliminates the need for phasing and 
the variations introduced by the use of separate antenna sets from epoch 
to epoch. Consequently, a large number of pulsars can be 
observed in an automatic mode. Moreover, such observations 
can be carried out concurrently with other GMRT observations 
exploiting the GMRT sub-array mode, where one antenna can be  used 
for pulsar observations concurrently with the other 29 observing 
a different (even a non-pulsar) source. 
Such a backend can be used with the existing 400 MHz wide 
21 cm feed on GMRT and will also provide a digital backend for 
other wide-band feeds proposed in the GMRT upgrade. The prime 
motivation of such an instrument is regular monitoring of pulsars 
for high precision timing studies. The feasibility of various designs 
is being evaluated currently.

\section{ACKNOWLEDGMENT}

We thank the staff of the GMRT for their help during tests. 
GMRT is run by the National Centre for Radio Astrophysics of 
the Tata Institute of Fundamental Research.

%\footnotesize
\bibliographystyle{basi}
%\bibliography{journals,journals_apj,modrefs,psrrefs,crossrefs}

\begin{thebibliography}{25}
\expandafter\ifx\csname natexlab\endcsname\relax\def\natexlab#1{#1}\fi
\expandafter\ifx\csname url\endcsname\relax
  \def\url#1{{\tt #1}}\fi
\expandafter\ifx\csname urlprefix\endcsname\relax\def\urlprefix{URL }\fi

\bibitem[\protect\citeauthoryear{{Deshpande}}{{Deshpande}}{1995}]{d95}
{Deshpande}, A.~A., 1995, {\em J. Astrophys. Astr.\/}, {\bf 16}, 225.

\bibitem[\protect\citeauthoryear{Foster \& Backer}{Foster \&
  Backer}{1990}]{fb90}
Foster, R.~S., Backer, D.~C., 1990, {\em ApJ\/}, {\bf 361}, 300.

\bibitem[\protect\citeauthoryear{{Hankins} \& {Rickett}}{{Hankins} \&
  {Rickett}}{1975}]{hr75}
{Hankins}, T.~H., {Rickett}, B.~J., 1975, in {\em Methods in Computational
  Physics Volume 14 --- Radio Astronomy\/}, New York: Academic Press.

\bibitem[\protect\citeauthoryear{{Hankins} et~al.}{{Hankins}
  et~al.}{2003}]{hkwe03}
{Hankins}, T.~H., et~al., 2003, {\em Nature\/}, {\bf 422}, 141.

\bibitem[\protect\citeauthoryear{Jenet et~al.}{Jenet et~al.}{1998}]{jak+98}
Jenet, F., et~al., 1998, {\em ApJ\/}, {\bf 498}, 365.

\bibitem[\protect\citeauthoryear{Joshi, Lyne, \& Kramer}{Joshi
  et~al.}{2003}]{jlk03}
Joshi, B.~C., Lyne, A.~G., Kramer, M., 2003, {\em Bull.\ Astr.\ Soc.\ India\/},
  {\bf 31}, 237.

\bibitem[\protect\citeauthoryear{{Joshi} et~al.}{{Joshi} et~al.}{2003}]{jlk+03}
{Joshi}, B.~C., et~al., 2003, in {\em Radio Pulsars\/}, eds. M.~Bailes, D.~J.
  Nice, S.~Thorsett, San Francisco: Astron.\ Soc.\
  Pac.\ Conf.\ Ser.\ Vol.\ {\bf 302}, pp. 321

\bibitem[\protect\citeauthoryear{Kinkhabwala \& Thorsett}{Kinkhabwala \&
  Thorsett}{2000}]{kt00}
Kinkhabwala, A., Thorsett, S.~E., 2000, {\em ApJ\/}, {\bf 535}, 365.

\bibitem[\protect\citeauthoryear{{Kramer}}{{Kramer}}{1998}]{kra98}
{Kramer}, M., 1998, {\em ApJ\/}, {\bf 509}, 856.

\bibitem[\protect\citeauthoryear{{Kramer}, {Johnston}, \& {van
  Straten}}{{Kramer} et~al.}{2002}]{kjv02}
{Kramer}, M., {Johnston}, S., {van Straten}, W., 2002, {\em MNRAS\/}, {\bf
  334}, 523.

\bibitem[\protect\citeauthoryear{Kramer et~al.}{Kramer et~al.}{1999}]{kll+99}
Kramer, M., et~al., 1999, {\em ApJ\/}, {\bf 526}, 957.

\bibitem[\protect\citeauthoryear{Lyne et~al.}{Lyne et~al.}{2004}]{lbk+04}
Lyne, A.~G., et~al., 2004, {\em Science\/}, {\bf 303}, 1153.

\bibitem[\protect\citeauthoryear{{Maron} et~al.}{{Maron} et~al.}{2000}]{mkk+00}
{Maron}, O., et~al., 2000, {\em A\&AS\/}, {\bf 147}, 195.

\bibitem[\protect\citeauthoryear{McKinnon \& Stinebring}{McKinnon \&
  Stinebring}{1998}]{ms98}
McKinnon, M., Stinebring, D., 1998, {\em ApJ\/}, {\bf 502}, 883.

\bibitem[\protect\citeauthoryear{{Ord} et~al.}{{Ord} et~al.}{2004}]{ovhb04}
{Ord}, S.~M., et~al., 2004, {\em MNRAS\/}, {\bf 352}, 804.

\bibitem[\protect\citeauthoryear{Popov et~al.}{Popov et~al.}{2002}]{pbc+02}
Popov, M.~V., et~al., 2002, {\em A\&A\/}, 171.

\bibitem[\protect\citeauthoryear{Rickett}{Rickett}{1996}]{r96}
Rickett, B., 1996, in {\em Pulsars - Problems and Progress (ASP Conf.~Ser.)\/},
  ed. M.~B. S.~Johnston, M. A.~Walker, San Francisco: Astron.\ Soc.\ Pac.\
  Conf.\ Ser.\ Vol.\ {\bf 105}, pp. 439.

\bibitem[\protect\citeauthoryear{Sallmen \& Backer}{Sallmen \&
  Backer}{1995}]{sb95}
Sallmen, S., Backer, D.~C., 1995, in {\em Millisecond Pulsars: A Decade of
  Surprise\/}, eds. A.~S. Fruchter, M.~Tavani, D.~C. Backer, Astron.\ Soc.\
  Pac.\ Conf.\ Ser.\ Vol.\ {\bf 72}, pp. 340.

\bibitem[\protect\citeauthoryear{Sirothia}{Sirothia}{2000}]{sir00}
Sirothia, S., 2000, {\em {Pulsar Observations using Phased Array of GMRT}\/},
  M. Sc. thesis, University of Pune.

\bibitem[\protect\citeauthoryear{Stairs, Thorsett, \& Camilo}{Stairs
  et~al.}{1999}]{stc99}
Stairs, I.~H., Thorsett, S.~E., Camilo, F., 1999, {\em ApJS\/}, {\bf 123}, 627.

\bibitem[\protect\citeauthoryear{{Subrahmanya} et~al.}{{Subrahmanya}
  et~al.}{1995}]{sdt+95}
{Subrahmanya}, C.~R., et~al., 1995, {\em J. Astrophys. Astr.\/}, {\bf 16}, 453.

\bibitem[\protect\citeauthoryear{{Swarup} et~al.}{{Swarup}
  et~al.}{1991}]{sak+91}
{Swarup}, G., et~al., 1991, {\em Curr. Sci.\/}, {\bf 60(2)}, 95.

\bibitem[\protect\citeauthoryear{Taylor \& Weisberg}{Taylor \&
  Weisberg}{1989}]{tw89}
Taylor, J.~H., Weisberg, J.~M., 1989, {\em ApJ\/}, {\bf 345}, 434.

\bibitem[\protect\citeauthoryear{van Straten et~al.}{van Straten
  et~al.}{2001}]{vbb+01}
van Straten, W., et~al., 2001, {\em Nature\/}, {\bf 412}, 158.

\bibitem[\protect\citeauthoryear{{Weisberg} \& {Taylor}}{{Weisberg} \&
  {Taylor}}{2002}]{wt02}
{Weisberg}, J.~M., {Taylor}, J.~H., 2002, {\em ApJ\/}, {\bf 576}, 942.

\end{thebibliography}

\label{lastpage}

\end{document}